\documentclass[conference]{templates/ieee/IEEEtran}
\IEEEoverridecommandlockouts

\usepackage[T1]{fontenc}
\usepackage[utf8]{inputenc}
\usepackage{amsmath,amssymb,amsfonts}
\usepackage{algorithmic}
\usepackage{graphicx}
\usepackage{textcomp}
\usepackage{xcolor}
\usepackage{url}
\usepackage[caption=false,font=footnotesize]{subfig}
\usepackage{multirow} 
\usepackage{array} 
\usepackage{placeins}
\usepackage{hyperref}
\usepackage[capitalise, noabbrev]{cleveref}

\usepackage{listings}

\lstdefinelanguage{CUDA}[ISO]{C++}{
  morekeywords={__global__, __device__, __host__, __shared__,
                __constant__, __managed__, __restrict__,
                threadIdx, blockIdx, blockDim, gridDim,
                warpSize, syncthreads},
}

\lstdefinestyle{cuda}{
  language=CUDA,
  basicstyle=\ttfamily\small,
  keywordstyle=\color{blue}\bfseries,
  commentstyle=\color{gray}\itshape,
  stringstyle=\color{green!40!black},
  showstringspaces=false,
  breaklines=true,
}

\lstdefinelanguage{PTX}{
  morekeywords={
    ld, st, mov, cvt, add, sub, mul, mad, fma, div,
    bra, call, ret, exit, bar, atom, red, tex,
    .global, .shared, .const, .local, .param,
    .reg, .pred, .f32, .u32, .s32, .b32, .v2, .v4, .v8
  },
  sensitive=true,
  alsoletter={.},
  morecomment=[l]{//},
  morestring=[b]",
}

\lstdefinestyle{ptx}{
  language=PTX,
  basicstyle=\ttfamily\small,
  breaklines=true,
}

\newcommand{\ptxinline}[1]{\lstinline[style=ptx]{#1}}

\newcommand{\thickhline}{\noalign{\hrule height 1.2pt}}

\begin{document}
\title{Optimizing Bloom Filters on Modern GPUs}

\author{
\IEEEauthorblockN{Daniel Jünger\IEEEauthorrefmark{1},
Kevin Kristensen\IEEEauthorrefmark{1},
Yunsong Wang\IEEEauthorrefmark{1},
Xiangyao Yu\IEEEauthorrefmark{2},
Bertil Schmidt\IEEEauthorrefmark{3}}
\IEEEauthorblockA{\IEEEauthorrefmark{1}NVIDIA Corporation,
\IEEEauthorrefmark{2}University of Wisconsin-Madison,
\IEEEauthorrefmark{3}Johannes Gutenberg University\\
\{\href{mailto:djuenger@nvidia.com}{djuenger}, \href{mailto:kkristensen@nvidia.com}{kkristensen}, \href{mailto:yunsongw@nvidia.com}{yunsongw}\}@nvidia.com,
\href{mailto:yxy@wisc.edu}{yxy@wisc.edu},
\href{mailto:bertil.schmidt@uni-mainz.de}{bertil.schmidt@uni-mainz.de}}
}

\maketitle

\begin{abstract}
Bloom filters are a fundamental data structure for approximate membership queries in applications ranging from analytics and databases to genomics. Deployed as prefilters, they eliminate irrelevant data before expensive downstream processing. As data-processing pipelines move onto GPUs, filtering must remain GPU-resident and keep pace with other stages. Although Bloom filters have been extensively optimized for CPUs, few implementations target GPUs, where fixed SIMD layouts map poorly to SIMT hardware and leave performance potential on the table. We present an architecture-aware GPU Bloom filter with tunable vectorization for performance portability across workloads, memory regimes, and GPU architectures. On NVIDIA B200, it sustains over $92\%$ of the measured random-access bound. At comparable false-positive rates, it outperforms the state-of-the-art GPU baseline by $15.4\times$ for lookup and $11.35\times$ for construction. These gains bring accurate Bloom filters to GPU-scale throughput previously reserved for high-error variants. The implementation is openly available in NVIDIA's cuCollections library~\cite{cucollections}: \url{https://github.com/NVIDIA/cuCollections}.
\end{abstract}


\begin{figure*}[t]
  \centering
  \includegraphics[width=\textwidth]{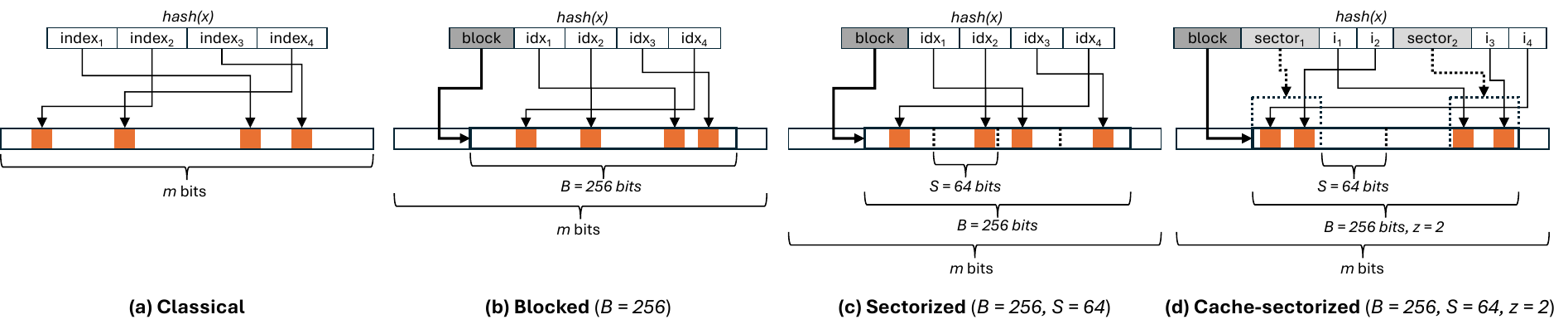}
  \caption{Overview of Bloom filter variants. Inspired by Schmidt et al.~\cite{schmidt2021}. For more details refer to \cref{sec:background:bloom}.}
  \label{fig:bf_variants}
\end{figure*}

\section{Introduction}
\label{sec:intro}

The rapid growth of data volumes poses challenges for large-scale analytics. Modern analytical systems increasingly execute complete data-processing pipelines directly on GPUs to sustain the required throughput~\cite{rapids2023,Aramburu2025,Yogatama2026}. Filtering eliminates irrelevant records early in the pipeline, reducing downstream memory traffic, communication, and computation~\cite{Shrinivas2013,Yang2024}. As more pipeline stages execute directly on GPUs, filtering must remain GPU-resident and keep pace with the surrounding processing; otherwise, its overhead can erase these savings.
Bloom filters~\cite{bloom1970} are space-efficient data structures for approximate membership queries. They trade exactness for storage efficiency: queries may report false positives but never false negatives. This asymmetric error property makes them suitable for prefiltering, where occasional unnecessary work is acceptable, but relevant records must not be discarded. Applications include large-scale analytics~\cite{patgiri2019}, network caching~\cite{broder2013}, database systems~\cite{gubner2019,lang2019}, and genomics~\cite{melsted2011,li2024,strannenheim2010,srikakulam2023,mehringer2023,pellow2017,mcvicar2017}.
Bloom filters have been extensively optimized for CPU architectures~\cite{schmidt2021,lang2019,putze2009,apple2021,polychroniou2014,polychroniou2015}, but comparatively few implementations target GPUs~\cite{gubner2019,juenger2020,hayashikawa2019}. Efficient GPU execution requires more than transferring fixed CPU SIMD-vectorization layouts to SIMT hardware. Existing GPU filter implementations either retain fixed execution mappings~\cite{juenger2020,ma2011} or prioritize additional functionality over construction and lookup throughput~\cite{geil2018,mccoy2023}. They therefore provide no general mechanism for adapting execution across workloads and GPU architectures.

To address this gap, we develop an architecture-aware GPU Bloom filter design with configurable execution mappings for performance portability. We evaluate the design across cache-resident and DRAM-resident filters and multiple GPU architectures. Our implementation is available as part of NVIDIA's cuCollections library~\cite{cucollections}. Its Apache Arrow/Parquet-compatible configuration powers production GPU-accelerated Parquet row-group pruning in NVIDIA RAPIDS cuDF~\cite{rapids2023,apple2021}.

\section{Background}
\label{sec:background}
\subsection{Bloom Filter}
\label{sec:background:bloom}
\subsubsection{Classical Bloom Filter (CBF)}
\label{sec:background:bloom_cbf}

A Classical Bloom filter (CBF)~\cite{bloom1970} (\cref{fig:bf_variants}(a)) approximates a set of $n$ distinct keys as a bit array of size $m$. During construction, $k$ hash functions map each key to $k$ positions in the array, which are set to one. During lookup, the same $k$ positions are recomputed and inspected. If any corresponding bit is zero, the key is certainly absent. If all $k$ bits are set, the filter reports membership, although this may be a false positive when each queried position was already set by one or more other keys. The bits per element, $c=\frac{m}{n}$, control the space-accuracy trade-off. \Cref{eq:cbf-relations} gives the expected false-positive rate (FPR) $f(c,k)$. For fixed $c$, it is minimized by $k^*=c\ln 2$, directly linking the accuracy-optimal hash count to the space budget.
\begin{equation}
    f(c,k) \approx \left(1-e^{-k/c}\right)^k,\qquad
    f(c,k^*)
    \approx\left(\tfrac{1}{2}\right)^{c\ln 2}.
\label{eq:cbf-relations}
\end{equation}

\subsubsection{Blocked Bloom Filter (BBF)}
\label{sec:background:bloom_bbf}

A CBF performs $k$ potentially distant accesses per key, leading to poor cache locality. Blocked Bloom Filters (BBFs)~\cite{putze2009} (\cref{fig:bf_variants}(b)) instead partition the array into $b$ blocks of $B=m/b$ bits and use an additional hash to select one block per key. All $k$ fingerprint bits then reside within the selected block, reducing each operation to one localized access. Assuming uniform block selection, each block acts as a smaller CBF holding approximately $n/b$ keys. Blocks are commonly sized near a cache line so one key requires only one cache-line access. The trade-off is a higher FPR than a same-size CBF because random block assignment produces uneven per-block occupancy. This loss can be mitigated by allocating additional space and retuning $k$.~\cite{putze2009}

\subsubsection{Register Blocked Bloom Filter (RBBF)}
\label{sec:background:bloom_rbbf}

An extreme instance of the BBF is the Register Blocked Bloom Filter (RBBF)~\cite{lang2019}, which sets $B$ to one machine word. All $k$ fingerprint positions are restricted to a single machine word, so construction requires a single atomic update, whereas lookup requires one load followed by a word-level comparison. Its small block, however, admits relatively few distinct $k$-bit patterns. Different keys therefore collide more frequently than in larger blocks, substantially increasing the FPR.

\subsubsection{Sectorized Bloom Filter (SBF)}
\label{sec:background:bloom_sbf}

The Sectorized Bloom Filter (SBF)~\cite{lang2019,apple2021} (\cref{fig:bf_variants}(c)) relaxes the one-word restriction of the RBBF by dividing a block into $s=B/S$ contiguous machine words of $S$ bits. It distributes the $k$ fingerprint bits evenly across these words, ideally assigning $k/s$ bits to each. This balances bit usage, reduces intra-block contention, and improves accuracy over an RBBF while retaining efficient word-level comparisons. Because the words are contiguous and can be processed independently, the layout supports vectorized loads and parallel word processing on CPUs and GPUs. Prior work calls these partitions \emph{sectors}; here, we call them \emph{words} to distinguish them from the CUDA memory sectors described in \cref{sec:background:gpumem}.

\subsubsection{Cache-Sectorized Bloom Filter (CSBF)}
\label{sec:background:bloom_csbf}

An SBF requires $k\geq s$, preferably with $k$ divisible by $s$, so every word receives the same number of fingerprint bits. This can force a high hash count even when the accuracy-optimal $k^*$ is smaller, increasing per-key computation. For example, $B=1024$ and $S=64$ imply $s=16$ and therefore $k\geq16$. Cache-Sectorized Bloom Filters (CSBFs)~\cite{lang2019} (\cref{fig:bf_variants}(d)) relax this constraint by partitioning the $s$ words into $z$ groups and selecting exactly one word per group. Each selected word receives $k/z$ fingerprint bits, so $k$ need only be divisible by $z$. Construction atomically updates and lookup reads only the $z$ selected words rather than all $s$, reducing memory traffic and the number of word-level patterns generated from $s$ to $z$. The trade-off is a runtime-dependent word-selection step that introduces address dependencies and nonuniform work.

\subsection{GPU Memory System}
\label{sec:background:gpumem}
Modern CUDA-enabled accelerators employ either GDDR or high-bandwidth memory (HBM), with HBM providing substantially higher bandwidth. Despite this bandwidth, global-memory access remains relatively high latency in both memory systems, making memory-level parallelism essential. The memory hierarchy consists of a two-level cache system: each Streaming Multiprocessor (SM) is equipped with a private L1 cache, while a unified L2 cache is shared across all SMs and also serves as the point of coherence for the CUDA memory architecture. The minimum L2 request granularity is 32 bytes, referred to as a sector, with four sectors composing a 128-byte cache line. Accesses from threads on the same SM that target the same cache line or even sector can be merged into a single L2 request through temporal coalescing. Consequently, coalesced memory access patterns are critical for the efficient use of GPU DRAM, as they minimize the number of high-latency memory transactions. Importantly, atomic updates benefit from the same coalescing mechanism as well, which we will show is essential for efficient concurrent, lock-free filter construction.

In the context of BBF and SBF, access patterns are typically irregular: each update or lookup of a key may target a different filter block, leading to semi-random memory accesses. Within a single filter block, accesses are contiguous and can often be coalesced into one or a few sector requests. However, when multiple blocks are processed concurrently by an SM, requests may be scattered across distant memory locations. More generally, the memory subsystem can only handle a limited number of concurrent requests. When many outstanding accesses accumulate, either because individual requests take long to complete or because too many are issued in parallel, contention arises. In this case, new requests cannot be serviced immediately, and the issuing warp must stall until resources free up. This makes the choice of filter block size critical, as it determines the trade-off between memory access efficiency, i.e., how well each sector transfer is utilized, and per-key computational cost, defined by how many words each thread must traverse to process a single key.

\section{Related Work}
\label{sec:related}

Early adaptations of Bloom filters for GPUs primarily focused on porting the classical variant (CBF) to the massive thread parallelism of graphics processors. Ma et al.~\cite{ma2011} and Shi et al. \cite{shi2010parallel} presented some of the first GPU implementations, utilizing global memory for the bit array. However, their approaches suffered from excessive uncoalesced memory accesses inherent to the CBF design.
More recent efforts have adopted blocked designs to exploit the GPU's cache hierarchy. WarpCore~\cite{juenger2020} provides a state-of-the-art BBF implementation. While efficient for specific configurations, WarpCore relies on a rigid thread-cooperation scheme that assigns a fixed number of threads to each key during both construction and lookup. Because construction and lookup have different execution characteristics that also change with block size, a fixed thread mapping may not always be optimal.
The Folded Bloom Filter~\cite{hayashikawa2019} improves the false-positive rate of BBFs on GPUs by distributing keys across paired bit arrays to equalize load. However, this technique necessitates complex address folding and hash generation logic that incurs computational overhead.
Gubner et al.~\cite{gubner2019} explored SBF designs on GPUs within a co-processing framework for accelerating database joins. While they demonstrated the viability of the SBF layout, their work focused on the system-level orchestration of data transfers between CPU and GPU rather than the micro-architectural optimization of the filter kernel itself.

Beyond Bloom filters, several other Approximate Membership Query (AMQ) data structures have been ported to GPUs, most notably Cuckoo filters~\cite{fan2014} and Quotient filters~\cite{geil2018,bender2012}.
Geil et al.~\cite{geil2018} presented a GPU implementation of Quotient filters, which support deletions and merging. However, their design relies on linear probing and complex metadata management, resulting in significantly lower update rates compared to Bloom filters. Incremental insertions slow down by $2.5\times$ compared to a standard GPU Bloom filter.
Similarly, GPU Cuckoo filters, as explored by McCoy et al.~\cite{mccoy2023} and Dortmann et al.~\cite{dortmann2026cuckoo}, offer higher space efficiency and deletion support. Yet, they suffer from stochastic control flow during construction (due to cuckoo kicking sequences) and dependent memory loads during lookup (checking alternate buckets). These effects serialize execution and make memory latency harder to hide than for independent, parallelizable Bloom-filter accesses.
While these structures are superior when deletion support is strictly required, they cannot match the raw throughput of an optimized SBF/CSBF for append-only or static set membership scenarios, which remain among the dominant use cases in large-scale analytics.


\section{Optimization Strategies}
\label{sec:impl}
In this work, we focus on an SBF design because its contiguous words can be processed independently, enabling vectorized memory access and thread-level parallelism on GPUs. Its block size parameter exposes a throughput–accuracy trade-off: smaller blocks reduce memory traffic and per-key work but increase the FPR, with the one-word RBBF at the fastest and least accurate extreme. Larger blocks improve accuracy but require more memory transactions and computation. We introduce three GPU-specific optimizations that sustain throughput as block size grows: configurable two-dimensional vectorization, branchless key-pattern generation, and work-efficient thread cooperation.

\subsection{Configurable Two-Dimensional Vectorization}
\label{sec:impl:vec}
We derive two complementary vectorization dimensions from the SBF layout described in \cref{sec:background:bloom_sbf}:
\begin{enumerate}
    \item[$\Phi$] \textbf{Vertical:} Words are contiguous in memory; $\Phi$ is the number of words processed by each thread per step.
    \item[$\Theta$] \textbf{Horizontal:} Words can be processed independently; $\Theta$ is the number of cooperating parallel threads assigned to one key.
\end{enumerate}

For filter lookups, property $\Phi$ enables vectorized memory accesses that fetch several neighboring words with one wide load. On older CUDA hardware, the widest load instruction available is 16 bytes (\ptxinline{ld.global.v4.u32} in the Parallel Thread Execution (PTX) ISA), corresponding to half a memory sector. Blackwell introduced 32-byte loads such as \ptxinline{ld.global.v8.u32}, which populate eight 32-bit registers spanning a full sector. A statically unrolled loop processes the loaded words and exposes independent per-word arithmetic while prefetching the next chunk, allowing the scheduler to overlap memory and computation.

Property $\Theta$ enables horizontal vectorization by distributing filter block words across multiple threads~\cite{polychroniou2015}. CUDA employs a Single-Instruction Multiple-Threads (SIMT) execution model, where threads are organized into warps of 32 that execute in lockstep. Warps can be partitioned into sub-warp Cooperative Groups (CGs), allowing several groups to process independent keys concurrently. Within this setup, a CG collectively processes the words of a filter block, with each thread responsible for a subset of the work. Register shuffles and warp votes provide low-latency communication without routing intermediate values through shared or global memory. For construction, a fully horizontal layout assigns neighboring words to converged threads within the same cooperative group so that their atomic instructions are issued nearly simultaneously, enabling the temporal coalescing mechanism discussed in \cref{sec:background:gpumem} to combine atomic updates to the same sector into a single L2 request.

\begin{figure}[t]
  \centering
  \includegraphics[width=\columnwidth]{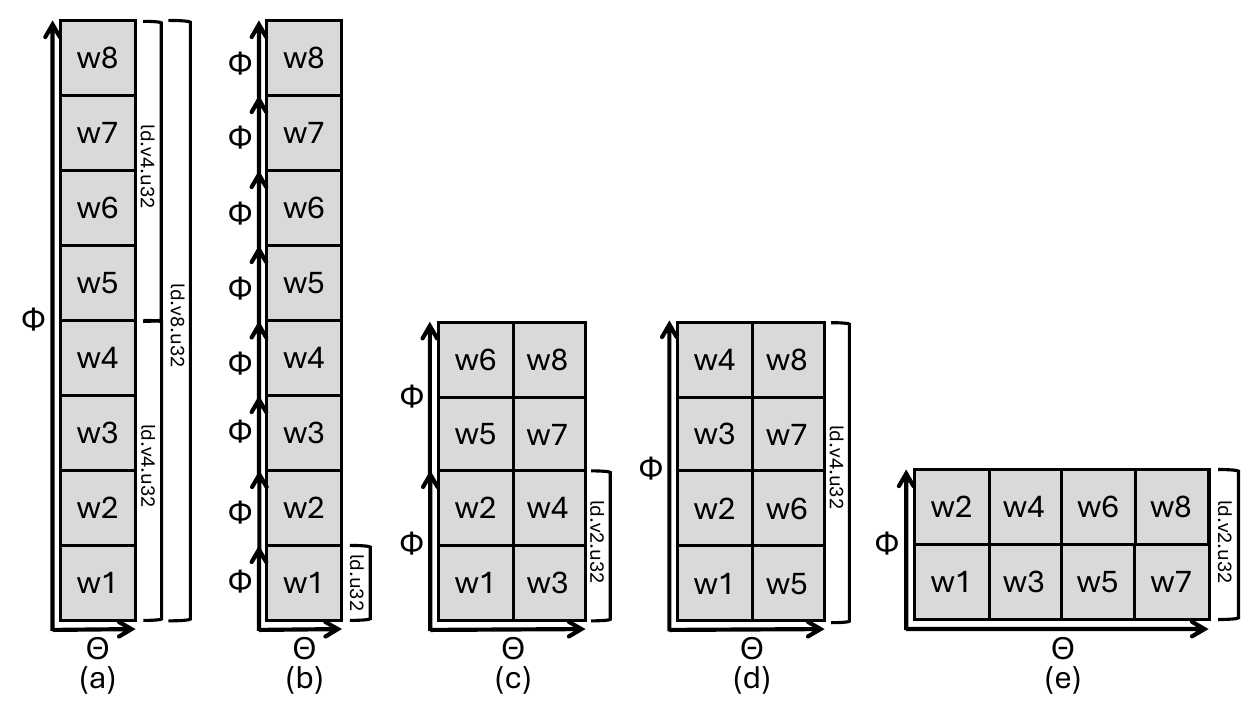}
  \caption{Representative vectorization layouts for $B=256$ and $S=32$ bits.
  Labels $w_1,\ldots,w_8$ show the order in which block words are assigned
  and processed, with execution progressing from bottom to top; right-side
  braces indicate memory load widths.}
  \label{fig:vec-layouts}
\end{figure}

For example, let $B=256$ and $S=32$, yielding $s=B/S=8$ words $w_1,\ldots,w_8$. \Cref{fig:vec-layouts} shows five representative layouts:
\begin{itemize}
    \item \textbf{(a): $\Theta=1,\Phi=8$}. One thread processes one key and loads the complete block with one 32-byte instruction on Blackwell+ or two 16-byte instructions on earlier GPU architectures. The eight words are then processed in an unrolled loop.
    \item \textbf{(b): $\Theta=1,\Phi=1$}. One thread processes one word at a time using scalar loads, preventing memory-access vectorization.
    \item \textbf{(c): $\Theta=2,\Phi=2$}. Two threads each load two contiguous words per step. The group traverses the block in strides of $\Theta\Phi=4$ words.
    \item \textbf{(d): $\Theta=2,\Phi=4$}. Each thread loads four words, covering the block in one step without a strided traversal.
    \item \textbf{(e): $\Theta=4,\Phi=2$}. Four threads each load two words, also covering the block in one step but distributing the work across more threads.
\end{itemize}

Valid layouts satisfy $1\leq\Theta\cdot\Phi\leq s$, with $\Theta$ and $\Phi$ restricted to powers of two. We enforce wide loads along $\Phi$ by representing each chunk as a statically sized, aligned array. This exposes the access width and alignment to the compiler, allowing it to emit the widest applicable instruction.

The two dimensions expose opposing resource trade-offs. Increasing $\Theta$ spreads one key across more threads, reducing per-thread work and register demand but requiring group communication and synchronization while leaving fewer independent filter operations in flight per warp. Favoring $\Phi$ over $\Theta$ assigns more work to each thread and allows more work items to execute concurrently per warp, while the resulting register demand from wide loads and unrolling may reduce occupancy or cause spills. In \cref{sec:eval}, we show that the preferred balance depends on the filter operation, block size, memory-residency regime, and GPU architecture.

\subsection{Branchless Key-Pattern Generation}
\label{sec:impl:key-pat}
An essential design aspect of Bloom filters is the generation of bit patterns from input keys. Typically, one or more hash functions are applied to each key to produce high-entropy hash bits that determine the bit positions to be set in the filter. A uniform distribution of these hash values minimizes collisions and thus improves accuracy. The conventional approach~\cite{bloom1970} employs $k$ independent hash functions, yielding excellent pattern diversity at a cost linear in $k$. Prior GPU implementations~\cite{juenger2020} compute one initial hash and derive subsequent values by repeatedly hashing the key with the previous value and a new seed; these $k$ sequential evaluations limit instruction- and thread-level parallelism. Double hashing~\cite{kirsch2008} reduces full hash evaluations to two and derives the remaining positions through linear combinations, but still requires per-position arithmetic.
Putze et al.~\cite{putze2009} generate fingerprints by looking up precomputed values in a small table. Although effective on some architectures, this scheme is inefficient on GPUs: random accesses to distinct constant-memory entries serialize, while staging the table in shared memory introduces bank conflicts as well as initialization and synchronization overhead. The resulting lookup cost can exceed that of arithmetic alternatives whose latency GPU parallelism can hide.

Multiplicative hashing generates key patterns by multiplying a base hash value with a set of odd constants (salts). The base hash value maps each key to a 64-bit value; our default implementation uses a high-throughput 64-bit implementation of xxHash~\cite{xxhash}, which provides excellent entropy characteristics. This technique has long been employed in CPU Bloom filter implementations~\cite{apple2021,polychroniou2014}. The base hash function is evaluated only once per key, and salted multiplications derive the remaining bit positions without a sequential dependency chain.
In vectorized CPU implementations, multiplicative hashing has been combined with both pure horizontal vectorization~\cite{apple2021} and pure vertical vectorization~\cite{polychroniou2014}. In both cases, the number of multipliers (salts) is fixed to the number of SIMD lanes so that they fit within a single SIMD register. While this approach is sufficient for CPUs, efficient GPU execution requires greater flexibility: both horizontal and vertical vectorization (see \cref{sec:impl:vec}) must be supported simultaneously, and the implementation must handle arbitrary combinations of $k$, $\Theta$, and $\Phi$.

Two challenges arise in this context:
\begin{enumerate}
  \item \textbf{Inlining salts into generated code.}
  A natural design is to place the multipliers in constant or shared memory for efficient reuse during key pattern generation. Both approaches require memory loads (\ptxinline{ld.const} or \ptxinline{ld.shared} in PTX) into registers, while staging the multipliers in shared memory additionally requires initialization and synchronization. To eliminate these loads and reduce access latency, the multipliers could instead be kept directly in registers. However, this scales poorly as $k$ increases because the resulting register pressure can reduce occupancy or cause local-memory spilling. We therefore inject the multipliers directly into the generated machine code. We statically unroll the inner loop over $\Phi$ words per thread and the outer loop over $\Theta \cdot \Phi$ word groups per cooperative group. Unrolling exposes as compile-time constants both the multiplier index, which allows the compiler to inline the corresponding multiplier literal, and the outer loop index so that the compiler can narrow the range of starting multiplier indices for the threads cooperating in pattern generation in an iteration of the inner loop.

  \item \textbf{Mapping thread indices to salt indices.}
  When $\Theta > 1$, threads in a cooperative group must select distinct starting multipliers. A naïve solution, such as a large switch statement, would generate excessive comparisons and long predicated instruction sequences. Instead, we construct a compile-time binary decision tree over the narrowed range of starting multiplier indices. This reduces the runtime selection depth to $\log(\Theta)$, ensuring minimal branching and predication overhead.
\end{enumerate}

Together, these techniques yield a mechanism that automatically specializes key pattern generation for arbitrary vectorization layouts, while maintaining inlined multipliers and efficient thread-to-salt mapping.

\subsection{Work-Efficient Thread Cooperation}
\label{sec:impl:adapt}
If $\Theta > 1$, the filter operation is executed in a group-cooperative manner, where two or more coalesced threads jointly process the same input key. This execution scheme requires that all participating threads operate on the same input data, which implies that computations independent of the group-local thread index yield the same result across the group. Such \emph{uniform} computations are therefore redundantly executed by every thread. While CUDA provides special registers and instructions for warp-uniform code paths, this support does not extend to independent sub-warp groups. Consequently, group-uniform work is replicated across the $\Theta$ cooperating threads. In the context of SBF, the computation of a key's hash value and corresponding filter block index are strictly group-uniform and thus redundantly evaluated under this execution model.

\begin{figure}[t]
  \centering
  \includegraphics[width=0.85\columnwidth]{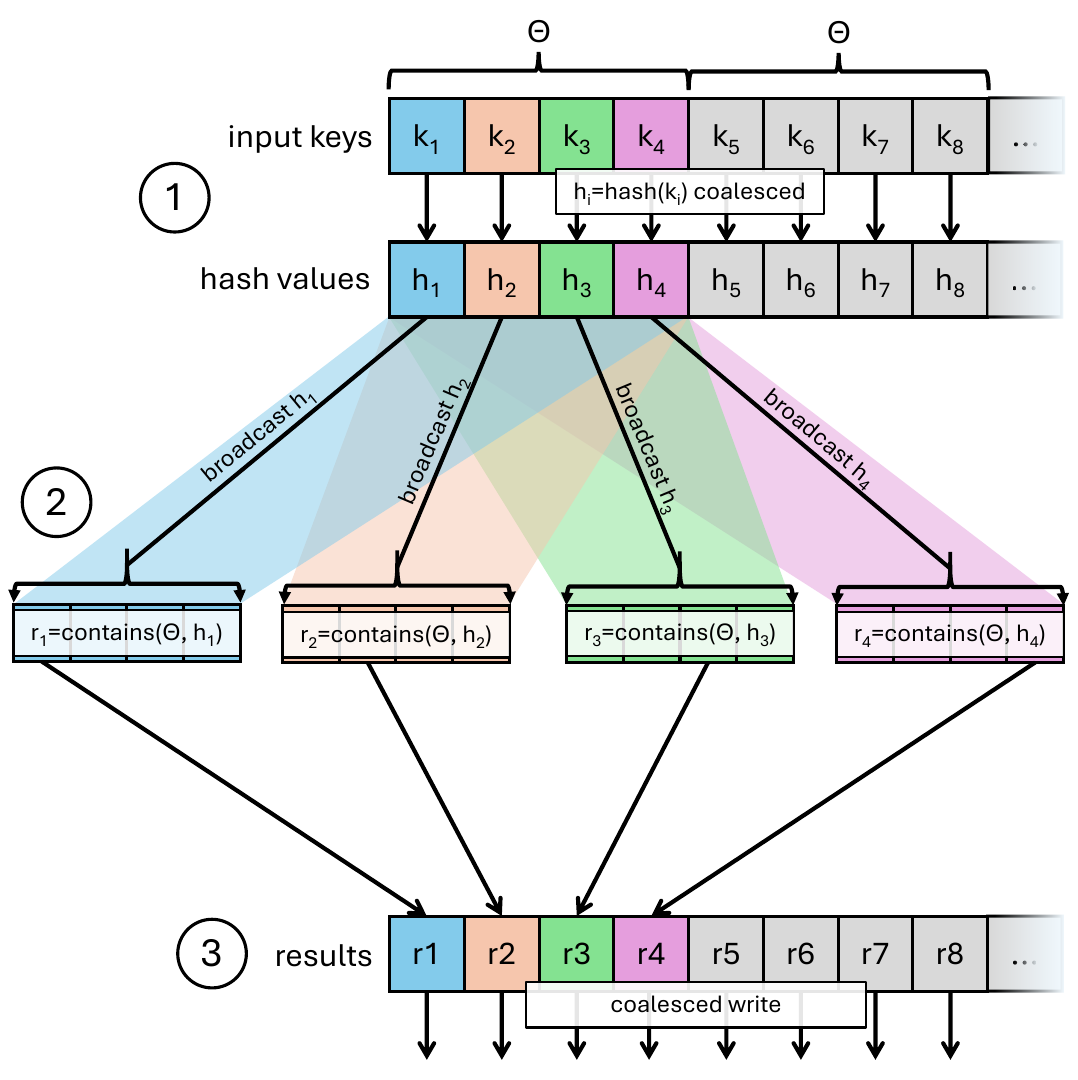}
  \caption{Work-efficient thread cooperation for bulk lookup with $\Theta=4$. Each lane first computes the hash for its assigned input key. The cooperative group then loops over the lanes, broadcasting one lane's hash at a time and performing the corresponding group-cooperative lookup. After keys $k_1,\ldots,k_4$ have been processed, each lane writes its own result, yielding coalesced output stores.}
  \label{fig:adapt-coop}
\end{figure}

To eliminate this redundant work, each thread first hashes a different key and later shares the resulting hash value and block index with its cooperative group, as illustrated in \cref{fig:adapt-coop}.
(1) A warp is assigned 32 consecutive input keys, creating a 1:1 mapping between threads and inputs. This ensures coalesced memory accesses and high cache-line utilization. Each thread computes the hash value and filter block index for its key and stores them in registers. For statically sized keys and hash functions without input-dependent control flow, all lanes execute this computation in lockstep.
(2) Each CG of size $\Theta$ then iterates over its lanes, broadcasting the hash value and block index of the currently active key via register shuffles. All threads in the group cooperatively execute the filter operation for that key.
(3) For lookup operations, each thread stores the result for its corresponding key in a register. After all group keys are processed, the results are written back in a coalesced fashion.

The evaluated configurations use $\Theta\leq16$ because the largest block contains only $s=16$ independent words. Larger groups bring no additional word-level parallelism while reducing the number of keys processed concurrently. Groups beyond one warp would also be unable to exchange values through register shuffles and would require shared-memory communication and additional synchronization.

\begin{figure*}[t!]
  \centering
  \makebox[\linewidth][c]{%
    \makebox[0.488\linewidth][c]{\scriptsize 32~MB filter (L2-resident)}\hspace{0.004\linewidth}
    \makebox[0.488\linewidth][c]{\scriptsize 1~GB filter (DRAM-resident)}%
  }\\[-0.2em]
  \makebox[\linewidth][c]{%
    \makebox[0.242\linewidth][c]{\hspace{0.01\linewidth}\scriptsize (a) construction}\hspace{0.004\linewidth}
    \makebox[0.242\linewidth][c]{\scriptsize (b) lookup}\hspace{0.004\linewidth}
    \makebox[0.242\linewidth][c]{\scriptsize (c) construction}\hspace{0.004\linewidth}
    \makebox[0.242\linewidth][c]{\scriptsize (d) lookup}%
  }\\[0.15em]
  \makebox[\linewidth][c]{%
    \llap{\raisebox{0.68in}{\rotatebox{90}{\scriptsize GElem/s}}\hspace{0.2em}}%
    \includegraphics[width=0.242\linewidth,trim=28pt 24pt 0 0,clip]{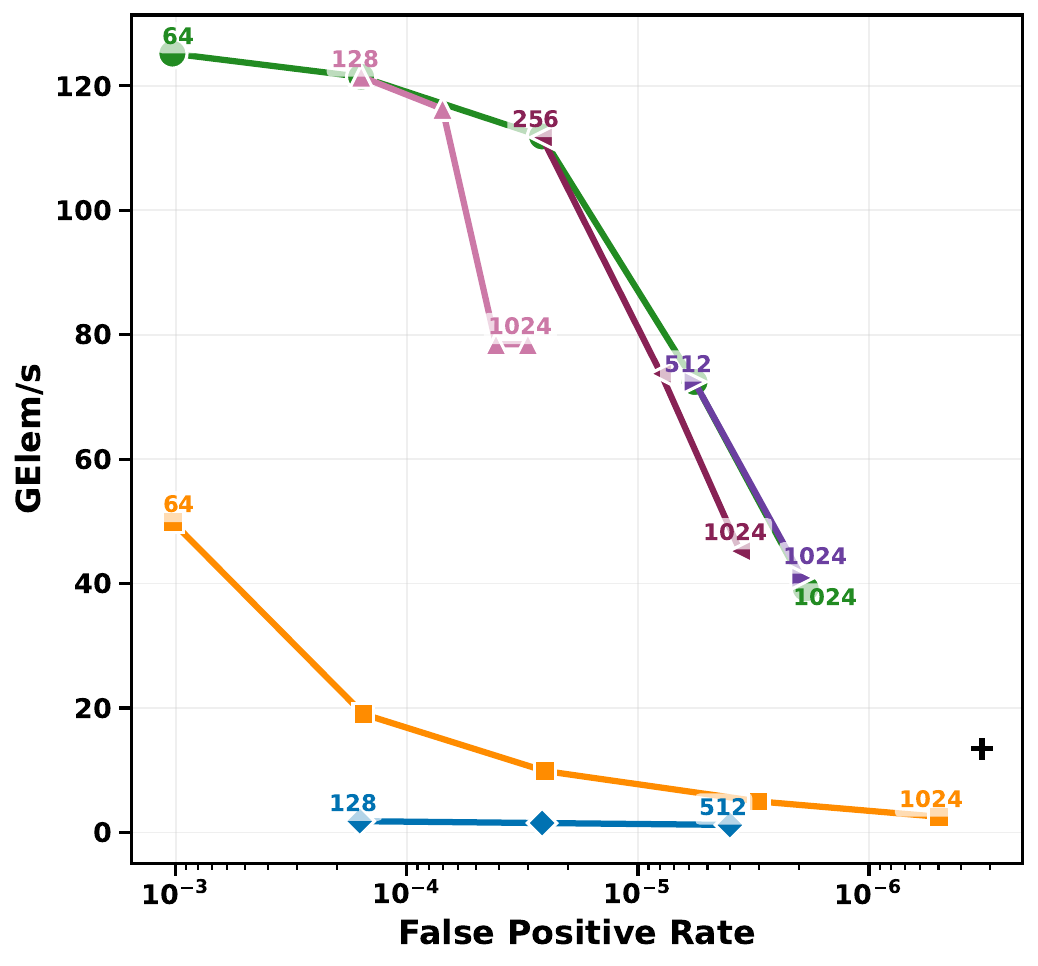}\hspace{0.004\linewidth}
    \includegraphics[width=0.242\linewidth,trim=28pt 24pt 0 0,clip]{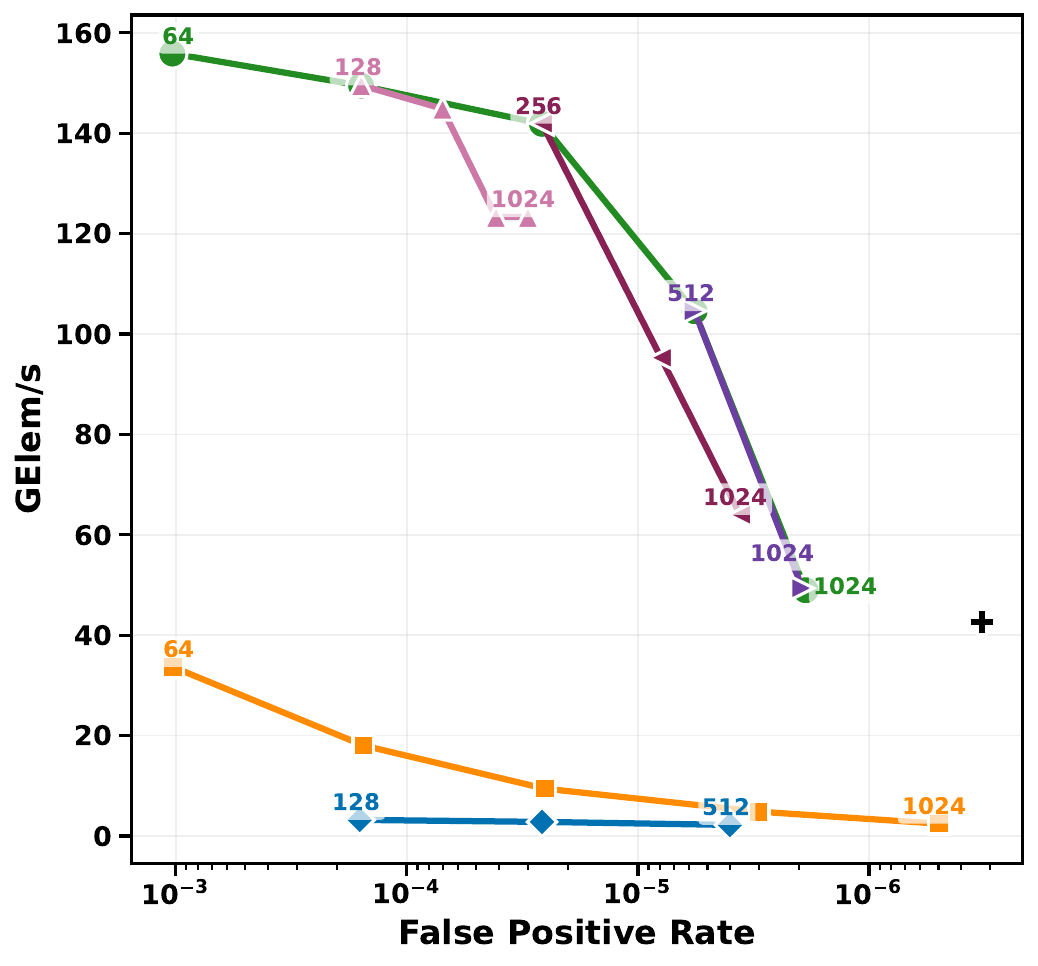}\hspace{0.004\linewidth}
    \includegraphics[width=0.242\linewidth,trim=28pt 24pt 0 0,clip]{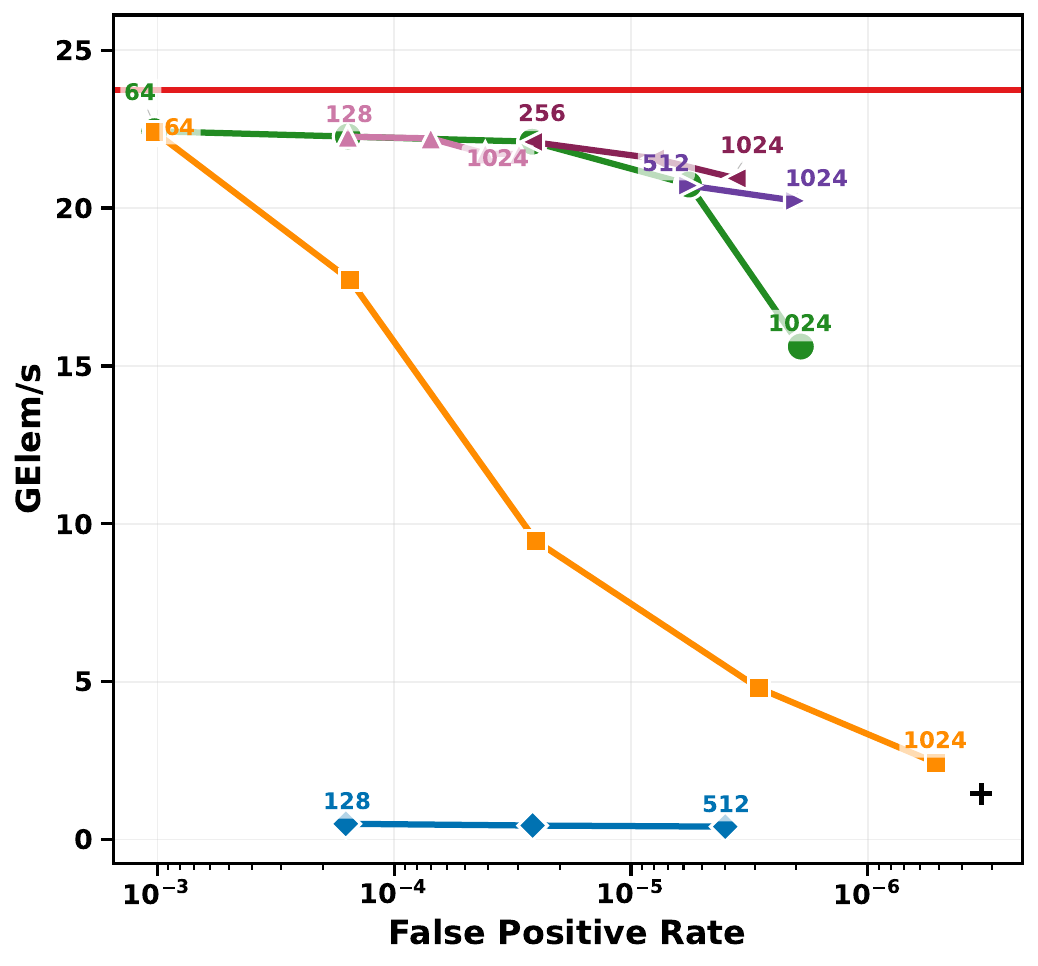}\hspace{0.004\linewidth}
    \includegraphics[width=0.242\linewidth,trim=28pt 24pt 0 0,clip]{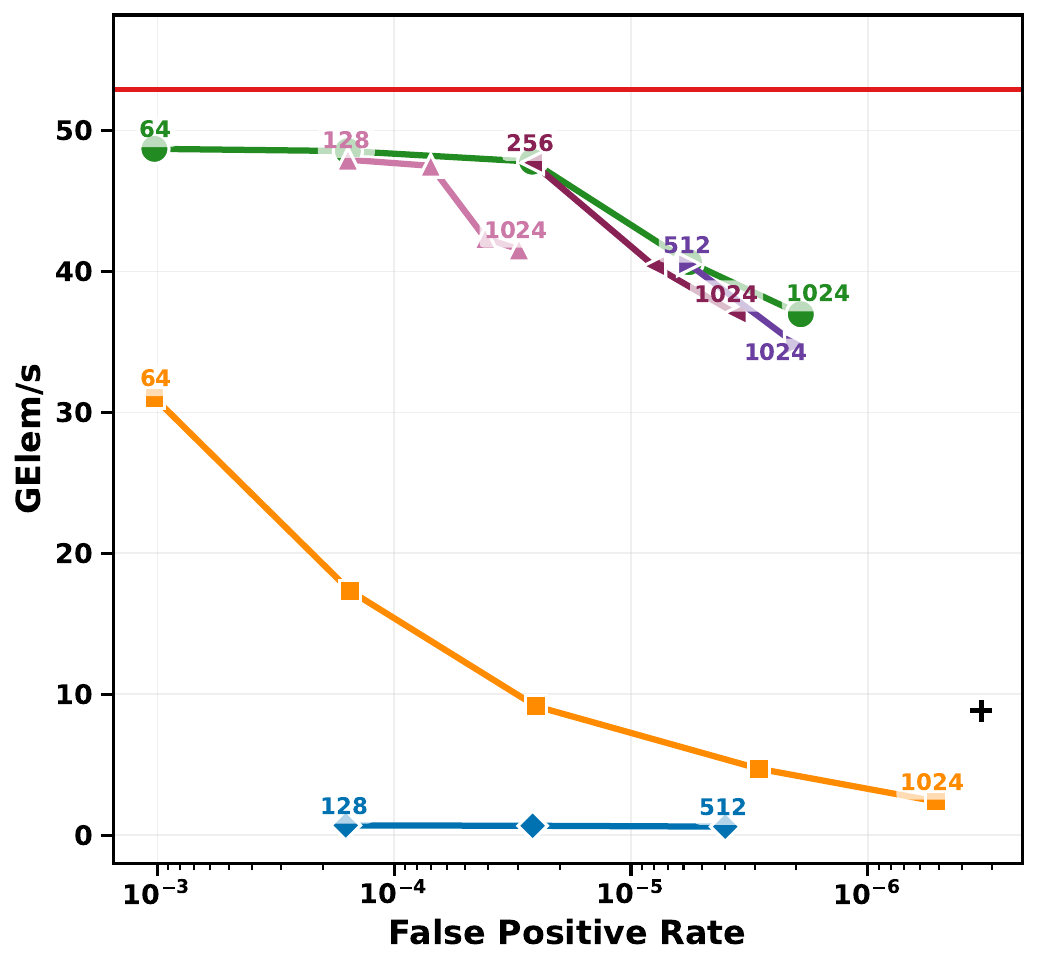}%
  }
  \\[-0.4em]
  \makebox[\linewidth][c]{%
    \makebox[0.242\linewidth][c]{\hspace{0.01\linewidth}\scriptsize False Positive Rate}\hspace{0.004\linewidth}
    \makebox[0.242\linewidth][c]{}\hspace{0.004\linewidth}
    \makebox[0.242\linewidth][c]{}\hspace{0.004\linewidth}
    \makebox[0.242\linewidth][c]{}%
  }\\[0.05em]
  \includegraphics[width=0.64\linewidth]{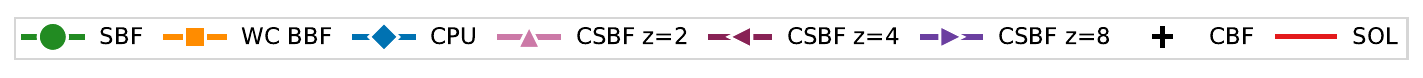}
\vspace{-0.4em}
  \caption{Throughput vs.\ false-positive rate frontier on NVIDIA B200 for construction and lookup with 32~MB (L2-resident) and 1~GB (DRAM-resident) filters. Data points are annotated with block size $B$ in bits; red lines mark the measured random-access throughput bound (SOL).}
  \label{fig:frontier}
\end{figure*}

\section{Experimental Evaluation}
\label{sec:eval}

\subsection{Evaluation Setup}
\label{sec:setup}
We evaluate performance and accuracy for L2-resident filters (\cref{sec:eval:cache}) and filters larger than cache (\cref{sec:eval:gmem}). Measured random-access throughput bounds quantify efficiency in the latter regime.
We compare against the GPU BBF from WarpCore~\cite{juenger2020} (WC BBF), the design previously used by NVIDIA's cuCollections library; a GPU CBF implemented in our framework; a GPU CSBF using the optimization techniques developed in \cref{sec:impl}; and a state-of-the-art multithreaded CPU SBF~\cite{schmidt2021}.
WC BBF uses a static, fully horizontal layout for both construction and lookup, corresponding to $(\Theta,\Phi)=(s,1)$.
In contrast, the CPU SBF uses a static, fully vertical layout in our terminology, with one CPU thread processing the eight words of each 512-bit block using AVX-512, corresponding to $(\Theta,\Phi)=(1,8)$, and radix-partitions large filters to keep random accesses within the CPU cache hierarchy.
For the GPU CSBF, we minimize inter-thread dependencies by requiring each group of $s/z$ words to fit within one vertical chunk, i.e., $s/z\leq\Phi$. The group index is calculated by introducing another odd multiplier to generate a new hash.

We run all experiments using NVIDIA's \textit{nvbench}~\cite{nvbench} benchmarking library, which performs multiple runs per configuration until measurement noise converges. We used CUDA 13.0.48, driver 580.65, and GCC 13.3.0. Our primary GPU is an NVIDIA B200; CPU results use a 16-core AMD EPYC 9124. The architecture comparison (\cref{sec:eval:arch}) additionally includes NVIDIA H200 SXM and NVIDIA RTX PRO 6000 GPUs, covering two memory systems: B200 and H200 use HBM3e, whereas RTX PRO 6000 uses GDDR7.

We set $S=64$ and $k=16$, where the latter is required to drive the maximum tested SBF block size of $B=1024$.
For each throughput measurement run, we generate $N=10^9$ unique, random $64$-bit input keys. To benchmark lookup, we pre-populate the filter with these keys, ensuring that all lookups yield true positive results.
The false-positive rate is measured by first inserting the space- and error-rate-optimal number of distinct keys into the filter, obtained from \cref{eq:cbf-relations}. We then query $N$ keys not present in the insertion set and record the fraction of false-positive responses.
Throughput and FPR are measured independently; the throughput benchmarks execute the same work regardless of filter occupancy.
\cref{fig:frontier} pairs each configuration's measured false-positive rate with its best construction and lookup throughput across all valid $(\Theta,\Phi)$ layouts; labels indicate $B$. In the construction panels, the false-positive rate characterizes the resulting filter, not the construction operation.
The architecture comparison likewise uses the best layout for each configuration. The tables instead compare $\Theta$ using the maximum valid $\Phi$, so each reported layout covers the entire block in one step.

\subsection{DRAM Filter}
\label{sec:eval:gmem}

\begin{table}[t]
\centering
\caption{Throughput (G Elem/s) of vectorization layouts for lookup and construction with a 1~GB filter on B200. Best layouts are bold. For each $\Theta$, we report only the maximum valid $\Phi$.}
\label{tab:layout_throughput_1024mb}
\setlength{\tabcolsep}{3.5pt} 

\begin{tabular}{|c|c!{\vrule width 1.2pt}c|c|c|c|c|}
\hline
\multirow{2}{*}{\textbf{Op.}} & \multirow{2}{*}{\boldmath{$B$}} & \multicolumn{5}{c|}{\boldmath{$\Theta$}} \\ \cline{3-7}
  & & \textbf{1} & \textbf{2} & \textbf{4} & \textbf{8} & \textbf{16} \\ \thickhline
\multirow{5}{*}{\rotatebox{90}{lookup}}
& \textbf{64}   & \textbf{48.69} & & & & \\ \cline{2-7}
& \textbf{128}  & \textbf{48.54} & 44.62 & & & \\ \cline{2-7}
& \textbf{256}  & \textbf{47.79} & 43.74 & 41.64 & & \\ \cline{2-7}
& \textbf{512}  & 25.35 & \textbf{40.66} & 40.15 & 33.66 & \\ \cline{2-7}
& \textbf{1024} & 12.81 & 36.01 & \textbf{36.96} & 33.38 & 24.54 \\ \thickhline
\multirow{5}{*}{\rotatebox{90}{construction}}
& \textbf{64}   & \textbf{22.43} & & & & \\ \cline{2-7}
& \textbf{128}  & 13.57 & \textbf{22.26} & & & \\ \cline{2-7}
& \textbf{256}  & 7.59  & 13.65 & \textbf{22.10} & & \\ \cline{2-7}
& \textbf{512}  &  4.58     & 7.72  & 15.31 & \textbf{20.75} & \\ \cline{2-7}
& \textbf{1024} &   2.88    &  5.02   & 8.53  & 15.41 & \textbf{15.61} \\ \hline
\end{tabular}
\end{table}

When the filter exceeds the GPU's L2 cache capacity, throughput is primarily constrained by DRAM random-access performance. We quantify this using random 64-bit load/store operations measured in giga-updates per second (GUPS)~\cite{DBLP:conf/sc/LuszczekBDKLRT06,nvidia_gups}. This setup closely resembles the access pattern of an RBBF and provides an upper bound for larger block sizes.

\cref{fig:frontier}(c) and (d) show that our SBF exceeds $92$\% of the measured random-access throughput bound for both construction and lookup when $B \leq 256$. In this DRAM-resident regime, reducing the block size below 256 bits does not yield additional performance gains, contrary to earlier assumptions that RBBFs ($B = 64$) offer the highest throughput. As discussed in \cref{sec:background:gpumem}, a block of at most 256 bits fits within one 32-byte sector request, so smaller blocks do not reduce the request count. On B200, bulk lookup favors $\Theta=1$ and $\Phi=s$ for $B\leq256$. For $B=512$ and $1024$, using $\Theta=2$ and $4$ is $1.6\times$ and $2.9\times$ faster than a fully vertical layout, respectively. It is beneficial to spread out the computation over multiple threads for larger filter block sizes, as computational work and register pressure (affecting occupancy) scale proportionally to $s$. However, when using more than one thread per work item, the additional communication and synchronization among CG threads incur further overhead, resulting in sublinear performance gains. The measured optima can be summarized as $\Theta=\max(1,\frac{B}{256})$, corresponding to one thread per sector.

Construction differs notably from lookup. On B200, a fully horizontal layout ($\Theta=s$) yields the highest measured construction throughput across all evaluated block sizes. Profiling indicates that construction throughput is dominated by the effectiveness of the L1 request coalescing mechanism, which merges per-SM accesses to the same cache line into a single request to L2. Increasing $\Theta$ from 1 to $s$ reduces the number of sector requests proportionally to $\tfrac{B}{S}$.
A fully horizontal layout exploits temporal coalescing by having converged threads issue all atomic updates to a block together, ideally within the same clock cycle, thereby minimizing interleaving with accesses to other cache lines and enabling the updates to be combined into the fewest possible L2 transactions.
Bulk operations generate many concurrent memory requests to hash-selected blocks throughout the filter. For $B > 256$, each key additionally requires multiple sector requests within its block, and the resulting pressure on memory request queues becomes performance-limiting. The profiler exposes this pressure differently for lookup and construction. Lookup remains active while issuing and consuming sector reads, so backpressure appears as \textit{stall\_mio} when the MIO instruction queue fills. Construction has no return-value dependency on its atomic updates, allowing warps to reach kernel exit while atomics remain pending; the same underlying pressure therefore appears primarily as \textit{stall\_drain}.

The frontiers for the GPU CSBF at larger block sizes exhibit the trade-off characteristic of this filter architecture: a smaller number of groups $z$ reduces memory traffic (benefiting memory-bound regimes) at the cost of a higher false-positive rate. However, the resulting performance gains are attenuated by the characteristics of the memory subsystem. \textit{First}, the 32-byte request granularity limits bandwidth savings. At $B=512$, the block spans two sectors, which all CSBF variants and the SBF access. At $B=1024$, $z=2$ accesses two sectors, whereas $z\geq4$ and the SBF access all four sectors of the cache line. \textit{Second}, the high latency of DRAM accesses often masks the reduction in transfer volume. This effect contrasts with cache-resident regimes (see \cref{sec:eval:cache} and \cref{fig:frontier}(a) and (b)), where the low latency of L2 access allows the reduced sector access count to translate more directly into throughput gains.

WC BBF approaches the random-access throughput bound for $B = 64$, but its performance declines rapidly as the block size increases. It employs a fully horizontal vectorization layout for both construction and lookup. Uneven per-word work limits update coalescing, while this fixed layout becomes increasingly suboptimal for lookup as $B$ grows (\cref{tab:layout_throughput_1024mb}).
The GPU CBF has the lowest measured FPR, roughly two orders of magnitude below our $B=256$ SBF, but sustains only $1.44$ and $8.66$~GElem/s for construction and lookup; our SBF is $15.3\times$ and $5.5\times$ faster, respectively.
The CPU SBF baseline sustains approximately $0.45$~GElem/s for construction and $0.65$~GElem/s for lookup.

\subsection{Cache-resident Filter}
\label{sec:eval:cache}

\begin{table}[h]
\centering
\caption{Throughput (G Elem/s) of vectorization layouts for lookup and construction with a 32~MB (L2-resident) filter on B200. Best layouts are bold. For each $\Theta$, we report only the maximum valid $\Phi$.}
\label{tab:layout_throughput_32mb}
\setlength{\tabcolsep}{3.5pt} 

\begin{tabular}{|c|c!{\vrule width 1.2pt}c|c|c|c|c|}
\hline 
\multirow{2}{*}{\textbf{Op.}} & \multirow{2}{*}{\boldmath{$B$}} & \multicolumn{5}{c|}{\boldmath{$\Theta$}} \\ \cline{3-7}
  & & \textbf{1} & \textbf{2} & \textbf{4} & \textbf{8} & \textbf{16} \\ \thickhline 
\multirow{5}{*}{\rotatebox{90}{lookup}}
& \textbf{64}   & \textbf{155.89} & & & & \\ \cline{2-7}
& \textbf{128}  & \textbf{149.50} & 51.58 & & & \\ \cline{2-7}
& \textbf{256}  & \textbf{141.88} & 51.57 & 50.40 & & \\ \cline{2-7}
& \textbf{512}  & \textbf{104.55} & 50.20 & 50.35 & 45.34 & \\ \cline{2-7}
& \textbf{1024} & 44.87 & \textbf{48.95} & 48.69 & 45.22 & 42.11 \\ \thickhline 
\multirow{5}{*}{\rotatebox{90}{construction}}
& \textbf{64}   & \textbf{125.19} & & & & \\ \cline{2-7}
& \textbf{128}  & 66.07 & \textbf{121.45} & & & \\ \cline{2-7}
& \textbf{256}  & 33.91 & 63.25 & \textbf{111.88} & & \\ \cline{2-7}
& \textbf{512}  & 17.10    & 20.67 & 35.56 & \textbf{72.41} & \\ \cline{2-7}
& \textbf{1024} & 8.19   & 10.37    & 11.55 & 18.91 & \textbf{39.22} \\ \hline 
\end{tabular}
\end{table}

When the filter fits entirely within the GPU’s L2 cache (\cref{fig:frontier}(a) and (b)), the workload is no longer dominated by DRAM random-access throughput. Instead, profiling reveals increasing utilization of the compute pipelines as the filter block size grows. In this cache-resident regime, the performance advantages of our SBF and CSBF implementations become substantially more pronounced. Across all tested configurations, our approach exhibits markedly lower compute-pipeline congestion than WC BBF, largely due to branchless key-pattern generation and, when $\Theta>1$, the elimination of redundant hash and block-index computations (\cref{sec:impl:key-pat,sec:impl:adapt}).
\cref{tab:layout_throughput_32mb} summarizes the throughput achieved by different vectorization layouts. As in the DRAM-resident case, a fully horizontal layout provides the highest performance for filter construction. For lookups, a purely vertical layout is substantially more effective if $B \leq 512$. With $\Theta = 1$, no inter-thread communication or synchronization is required. Once $\Theta > 1$, cooperative groups must exchange data and synchronize, introducing overhead that outweighs the benefits of horizontal parallelism for these block sizes.

For the GPU CSBF, the lower latency of L2 exposes sector traffic more directly. At $B=512$, all CSBF variants and the SBF access two sectors; at $B=1024$, $z=2$ accesses two sectors while $z\geq4$ and the SBF access four. Though more compute-bound in this regime, the CSBF and SBF exhibit similar instruction counts for a given block size, and the effect of the runtime-dependent group index calculation for the CSBF is invisible, as the compiler can freely reorder the index calculation and the lookup due to the static loop unrolling. With approximate parity in computation, the reduced sector count for $z=2$ at $B=1024$ becomes the deciding factor, producing a proportional improvement in throughput.

For the RBBF configuration ($B = 64$), our SBF achieves a $2.51\times$ ($4.63\times$) construction (lookup) speedup over WC BBF, increasing to $11.35\times$ ($15.4\times$) for $B = 256$.
The GPU CBF achieves $12.53$~GElem/s for construction and $42.61$~GElem/s for lookup. Its lookup throughput exceeds WC BBF across all tested configurations, pointing to fundamental inefficiencies in WarpCore’s kernel for small, cache-resident filters.
Finally, a CPU SBF on a modern 16-core processor achieves $1.2$ ($8.8$)~GElem/s for construction (lookup).

\subsection{GPU Architecture Comparison}
\label{sec:eval:arch}

\begin{figure*}[t]
  \centering
  \makebox[\linewidth][c]{%
    \makebox[0.488\linewidth][c]{\scriptsize 32~MB filter (L2-resident)}\hspace{0.004\linewidth}
    \makebox[0.488\linewidth][c]{\scriptsize 1~GB filter (DRAM-resident)}%
  }\\[-0.2em]
  \makebox[\linewidth][c]{%
    \makebox[0.242\linewidth][c]{\scriptsize (a) construction}\hspace{0.004\linewidth}
    \makebox[0.242\linewidth][c]{\scriptsize (b) lookup}\hspace{0.004\linewidth}
    \makebox[0.242\linewidth][c]{\scriptsize (c) construction}\hspace{0.004\linewidth}
    \makebox[0.242\linewidth][c]{\scriptsize (d) lookup}%
  }\\[0.15em]
  \makebox[\linewidth][c]{%
    \llap{\raisebox{0.62in}{\rotatebox{90}{\scriptsize GElem/s}}\hspace{0.2em}}%
    \includegraphics[width=0.242\linewidth,trim=28pt 28pt 0 0,clip]{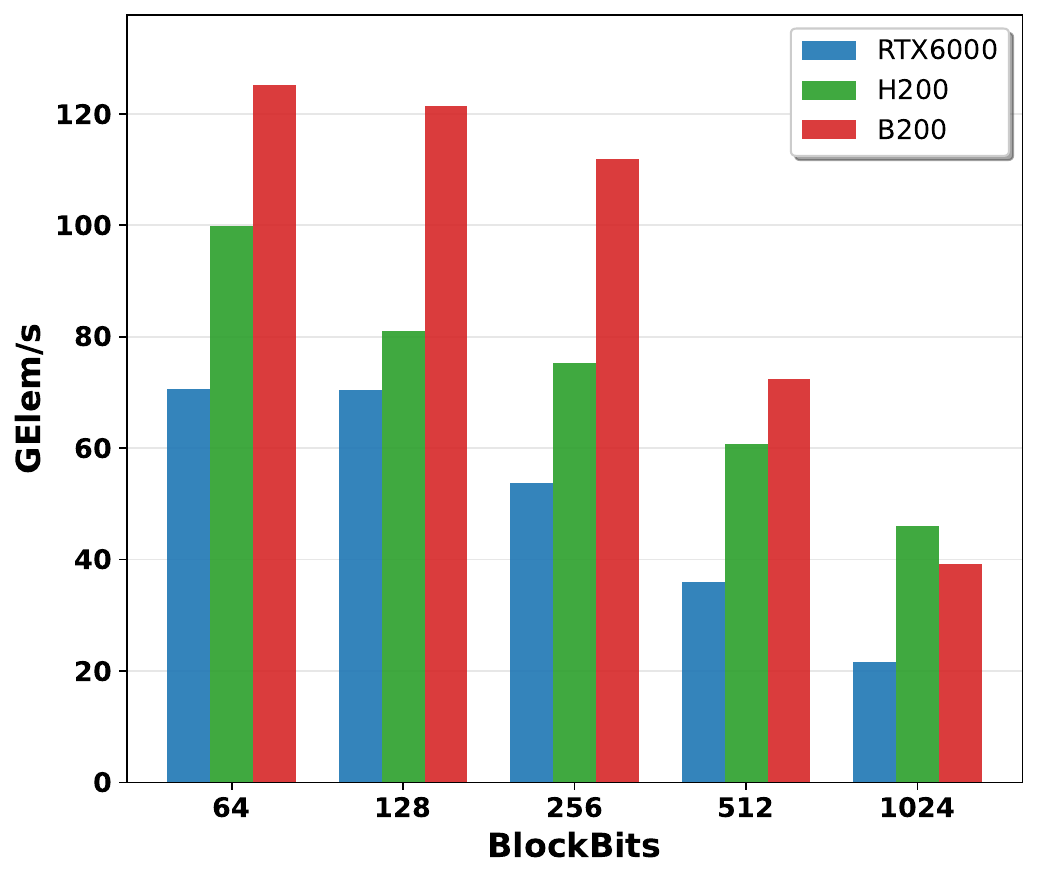}\hspace{0.004\linewidth}
    \includegraphics[width=0.242\linewidth,trim=28pt 28pt 0 0,clip]{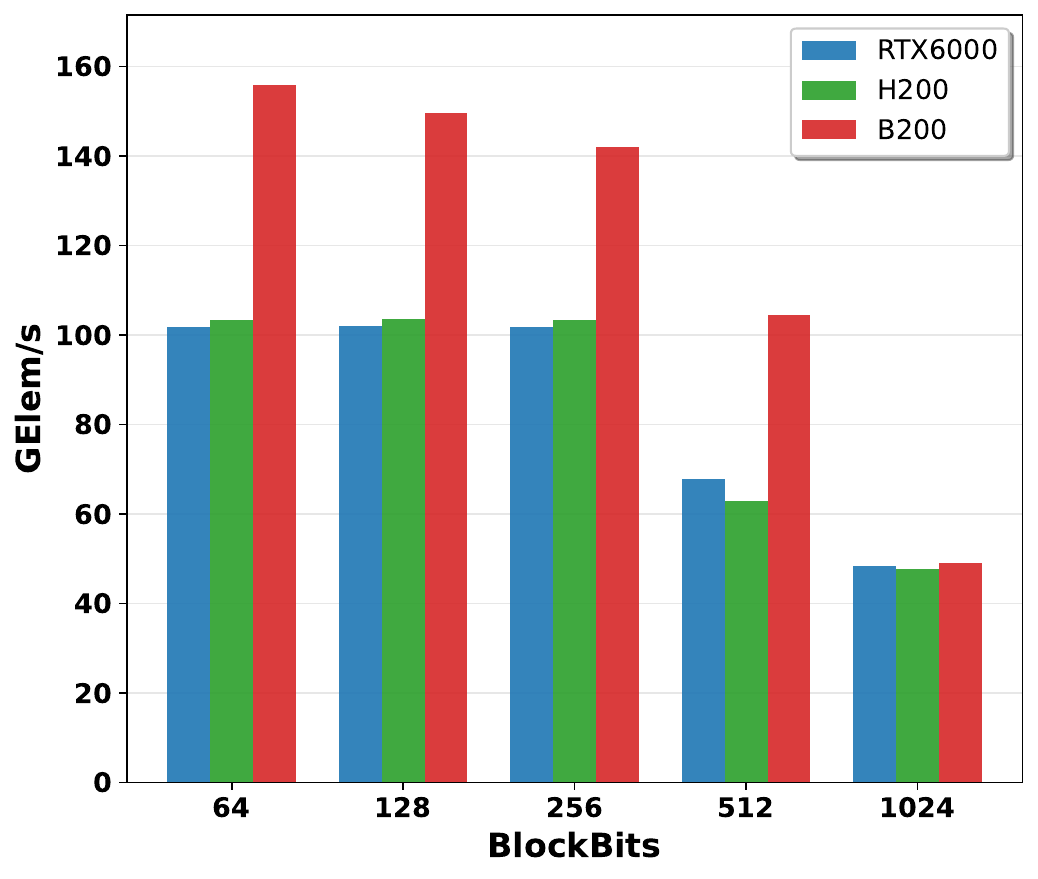}\hspace{0.004\linewidth}
    \includegraphics[width=0.242\linewidth,trim=28pt 28pt 0 0,clip]{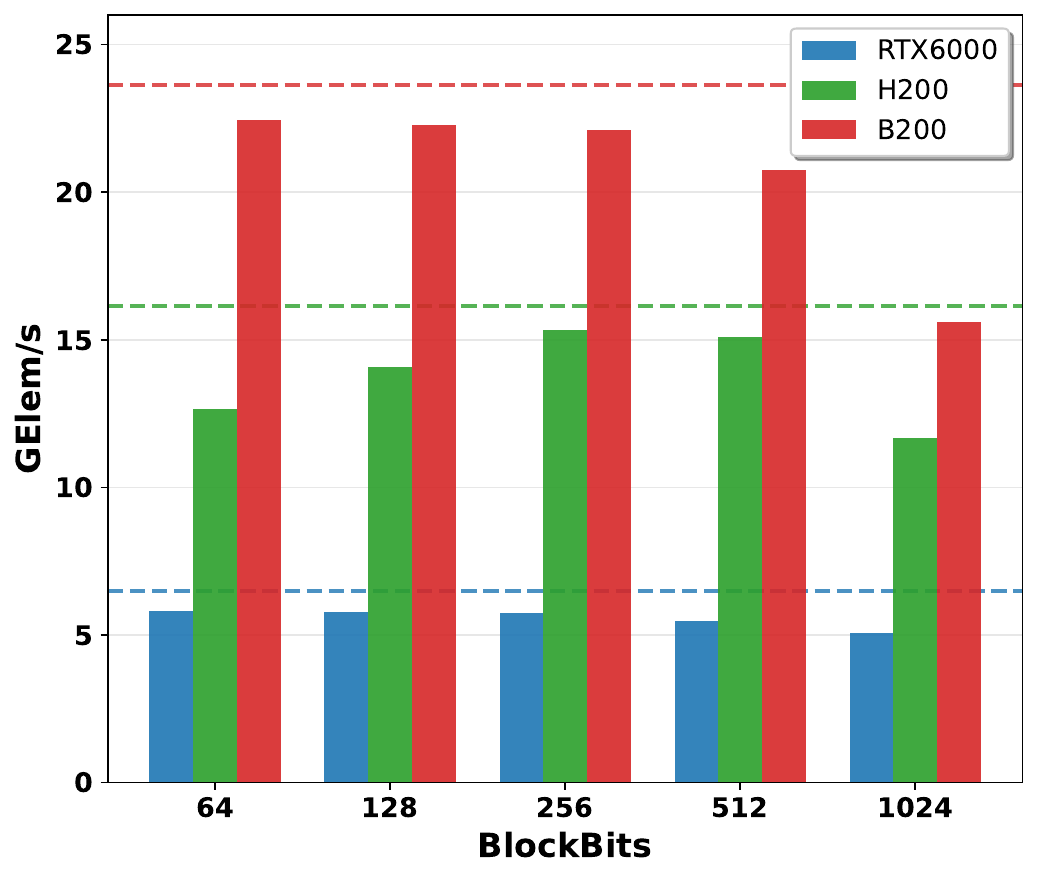}\hspace{0.004\linewidth}
    \includegraphics[width=0.242\linewidth,trim=28pt 28pt 0 0,clip]{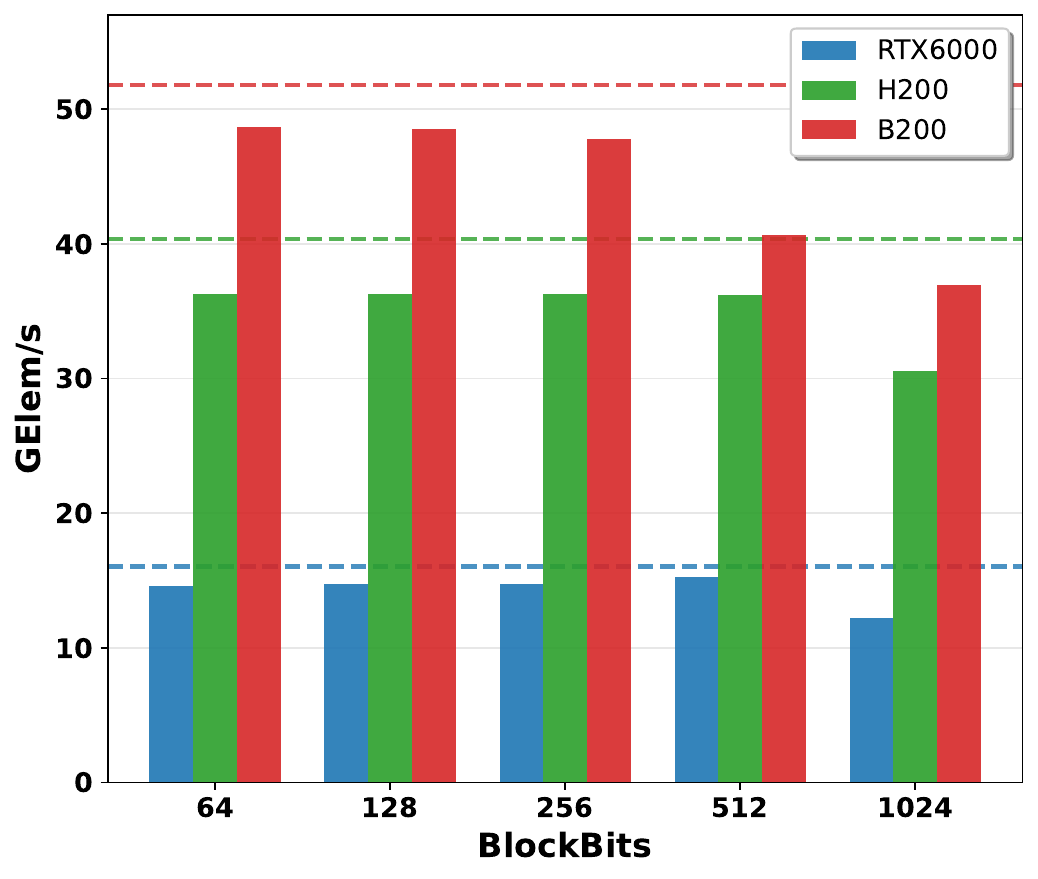}%
  }
  \\[-0.25em]
  \makebox[\linewidth][c]{%
    \makebox[0.242\linewidth][c]{\hspace{0.01\linewidth}\scriptsize BlockBits}\hspace{0.004\linewidth}
    \makebox[0.242\linewidth][c]{}\hspace{0.004\linewidth}
    \makebox[0.242\linewidth][c]{}\hspace{0.004\linewidth}
    \makebox[0.242\linewidth][c]{}%
  }
  \caption{Bulk construction and lookup throughput for 32~MB and 1~GB SBF filters across GPU architectures. Dashed lines in the 1~GB panels show the measured random-access throughput bound for each architecture.}
  \label{fig:arch-comparison}
\end{figure*}

\Cref{fig:arch-comparison} demonstrates the performance portability enabled by our configurable two-dimensional vectorization. For DRAM-resident filters, throughput tracks each GPU's measured random-access bound, reaching ${\sim}90$--$95\%$ for both operations across the HBM3e-based B200 and H200 and the GDDR7-based RTX PRO 6000. For L2-resident filters, RTX PRO 6000 remains competitive with H200 despite its lower DRAM bandwidth because its 188 SMs and higher clock rate provide more compute throughput than H200's 132 SMs.
Crucially, the optimal layout varies across architectures. For L2-resident construction at $B=512$ and $1024$, H200 favors $\Theta=4$ and $8$, whereas B200 and RTX PRO 6000 favor $\Theta=8$ and $16$, respectively. For DRAM-resident lookup at $B=128$ and $256$, RTX PRO 6000 favors $\Theta=2$, while B200 and H200 favor $\Theta=1$. These differences show why a fixed layout cannot provide performance portability across GPU architectures and memory regimes.

\subsection{Optimization Breakdown}
\label{sec:eval:opt}

\begin{figure}[t]
  \centering
  \includegraphics[width=\columnwidth]{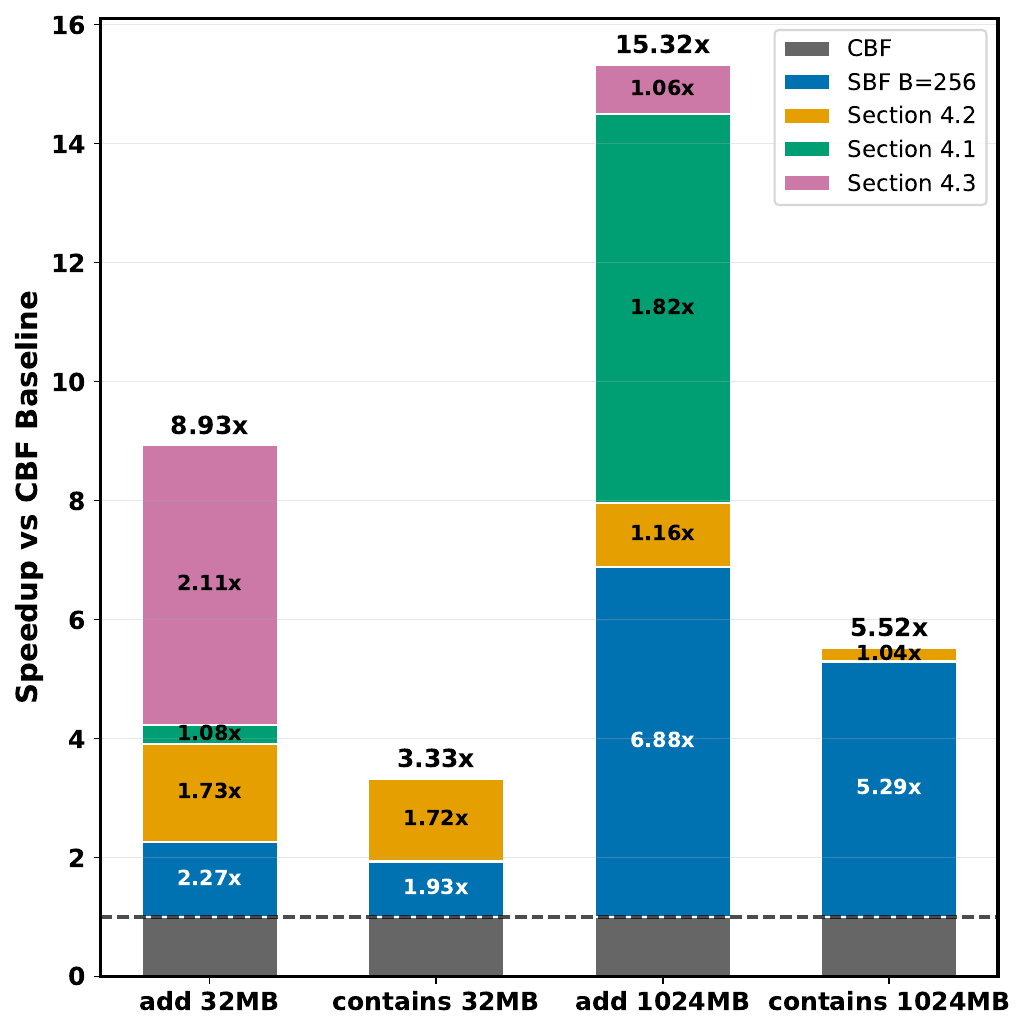}
  \caption{Relative performance breakdown comparing our optimized SBF with a GPU CBF baseline and an unoptimized SBF ($B=256$).}
  \label{fig:rel-speedup}
\end{figure}

\Cref{fig:rel-speedup} shows the sequential transition from a GPU CBF to an optimized SBF ($B=256$). The SBF layout reduces random sector accesses, particularly for DRAM-resident filters, while branchless multiplicative hashing yields a $1.72\times$ cache-resident speedup over the SBF baseline. Construction additionally benefits from horizontal vectorization and sharing hashes and block indices among cooperating threads, which together provide up to $2\times$: coalescing dominates in DRAM, whereas eliminating redundant computation matters more in L2. Lookup retains $\Theta=1$, so these cooperative optimizations do not apply.

\section{Conclusion}
\label{sec:conclusions}

We presented an architecture-aware GPU Bloom filter that combines configurable two-dimensional vectorization, branchless key-pattern generation, and work-efficient thread cooperation.
Our configurable design supports tuning across workloads and GPU architectures, enabling performance portability.
On NVIDIA B200, our 256-bit SBF exceeds $92\%$ of the measured random-access bound, matching the DRAM-resident throughput of high-error one-word RBBFs while providing substantially lower false-positive rates.
At comparable error rates, the implementation outperforms WarpCore by up to $15.4\times$ for lookup and $11.35\times$ for construction.
Together, these results make accurate Bloom filters practical as prefilters that keep pace with GPU-resident data-processing pipelines.
A configuration bitwise compatible with Apache Arrow/Parquet Split-Block Bloom Filters~\cite{apple2021} powers GPU Parquet row-group pruning in RAPIDS cuDF~\cite{rapids2023}.
The implementation is openly available under the Apache 2.0 license as part of NVIDIA's cuCollections library~\cite{cucollections}.

\FloatBarrier

\bibliographystyle{templates/ieee/IEEEtran}
\bibliography{references}

\end{document}